# Shell-model treatment of the $\beta$ decay of $^{99}$Tc

M. Ramalho($^1$) and J. Suhonen($^1$)($^2$)

($^1$) *Department of Physics, University of Jyväskylä, P.O. Box 35, FI-40014, Jyväskylä, Finland*

($^2$) *International Centre for Advanced Training and Research in Physics (CIFRA), P.O. Box MG12, 077125 Bucharest-Magurele, Romania*

**Summary.** — In the present paper we treat the second-forbidden non-unique (2nd-nu) ground-state-to-ground-state $\beta^-$ decay $^{99}$Tc($9/2^+$) → $^{99}$Ru($5/2^+$), with a 100% branching ratio, within the framework of the nuclear shell model (NSM). The energy spectrum of the electrons emitted in this $\beta$-decay transition ($\beta$-electron spectrum) is sensitive to the wave functions of the involved initial ($9/2^+$) and final ($5/2^+$) nuclear ground states through the many involved nuclear matrix elements (NME). The $\beta$-electron spectrum of this transition is potentially indicative of the effective value, $g_A^{\rm eff}$, of the weak axial coupling, $g_A$, of crucial importance for extraction of information on beyond-the-standard-model physics from the results of the present and future rare-events experiments. We describe the $\beta$ spectral shape of this decay by using a state-of-the-art $\beta$-decay formalism and compute the many involved NME using the well established NSM Hamiltonians *jj45pnb* and *glekpn*. We have found a strong dependence of the $\beta$ spectral shape on the value of $g_A$ making it a good candidate for determination of the value of $g_A^{\rm eff}$ through comparison with the corresponding experimental $\beta$ spectral shape. We have also found an interesting dependence of the $\beta$ spectral shape on the value of the so-called small relativistic vector NME, sNME, used to match the computed half-life with the measured one.

## 1. – Introduction

Rare-events experiments typically look for beyond-the-standard-model (BSM) physics by, e.g., measuring rare nuclear $\beta$ decays, double $\beta$ decays, and (anti)neutrino and WIMP (weakly interacting massive particle) scatterings off nuclei in searches for astrophysical neutrinos and cold dark matter of the Universe. In recent years, a booming interest in these experiments has concentrated on studies of $\beta$ electrons (electrons emitted in $\beta^-$ decays) and their energy distributions, the so-called $\beta$-electron spectra. Experimental and/or theoretical information on these spectra is crucial for, e.g., resolving the anomalies related to the antineutrino flux from nuclear reactors and for resolving the common backgrounds in the rare-events experiments themselves. Also, pinning down the effective values of weak couplings is a considerable incentive for those present and future $\beta$-decay experiments able to tackle the spectral shapes of $\beta$ electrons, see the review [1].

Nuclear $\beta$ decays vary in complexity: from allowed to highly forbidden ones [2, 3]. Like in the case of allowed $\beta$ decays, also in the case of forbidden unique $\beta$ decays the lepton phase space can be separated from the nuclear part, resulting in a universal $\beta$ spectral shape, independent of nuclear-structure details [2]. Of special interest for the rare-events experiments are the forbidden non-unique $\beta$ decays for which the $\beta$ spectral shapes can be strongly nuclear-structure dependent through several nuclear matrix elements (NME). The values of these NME are non-trivially determined by the wave functions of the initial and final states of a $\beta$-decay transition.

In addition to the many NME, the (partial) half-life of a forbidden non-unique $\beta$ transition depends on the values of the weak vector and axial-vector couplings, $g_V$ and $g_A$ [3]. In atomic nuclei, the CVC (conserved vector current) hypothesis sets the value $g_V = 1.0$, whereas the PCAC (partially conserved axial-vector current) hypothesis leads only to the so-called effective value of $g_A$, recently discussed in the reviews [1, 4, 5]. On the other hand, the so-called bare-nucleon value $g_A = 1.27$ stems from data on the decay of an isolated neutron. Typically, the effective value of $g_A$ is quenched relative to the bare-nucleon value which can have drastic effects on the sensitivity estimates of rare-events experiments trying to detect the neutrinoless double beta decay [6, 7], of crucial importance in the search for the BSM physics. Only in rare cases there is an enhancement of $g_A$ present [8].

As discussed in [9, 10], information on the value of $g_A$ can be gained by using the so-called spectrum-shape method (SSM). Use of SSM requires a $\beta$-electron spectrum with a notable $g_A$ dependence in its shape. In this case information on the effective value of $g_A$ can be gained through the comparison of computed template $\beta$ spectra, for different $g_A$ values, with the measured one in the aim to find a match. Such SSM analyses of $\beta$-spectral shapes of individual $\beta^-$ transitions have been done recently for the fourth-forbidden non-unique $\beta$ decays of $^{113}$Cd and $^{115}$In in [11, 12, 13].

An additional ingredient in the theoretical analyses of $\beta$ spectral shapes is the so-called small relativistic vector NME, the sNME, used to fix the measured (partial) half-life in the so-called enhanced SSM [14, 15]. In spite of its smallness, sNME can influence the (partial) half-lives and shapes of $\beta$-electron spectra quite strongly, see [12, 16, 17]. The sNME gathers its major contributions outside the proton (neutron) valence major shell that contains the proton (neutron) Fermi surface making it hard to be calculated in many nuclear models.

In the present paper we investigate the $g_A$ and sNME dependence of the second-forbidden non-unique (2nd-nu) $\beta^-$ decay of the $9/2^+$ ground state of $^{99}$Tc to the $5/2^+$ ground state of $^{99}$Ru, with a branching ratio of 100%.

## 2. – Theoretical Background

**2**`.`1. *NSM Hamiltonians.* – The NSM calculations were performed using the software KSHELL [18] with the interactions *jj45pnb* [19] and *glekpn* [20] with the set of single-particle energies fit for the mass region of $A = 94 - 98$. Their model spaces consist of the proton $0f_{5/2}(^2)$-$1p_{3/2}$-$1p_{1/2}$-$1g_{9/2}$ orbitals and the neutron $0g_{7/2}$-$1d_{5/2}$-$1d_{3/2}$-$2s_{1/2}$-$0h_{11/2}$ orbitals for *jj45pnb*, as well as $0f_{7/2}(^1)$-$0f_{5/2}(^2)$-$1p_{3/2}$-$1p_{1/2}$-$1g_{9/2}$ proton orbitals and $0g_{9/2}(^1)$-$0g_{7/2}$-$1d_{5/2}$-$1d_{3/2}$-$2s_{1/2}$ neutron orbitals for *glekpn*. The restrictions in the orbital occupations are imposed in order to reduce the computational burden. We believe, however, that these posed limitations do not affect much the wave functions of the initial and final ground states, relevant for the present work.

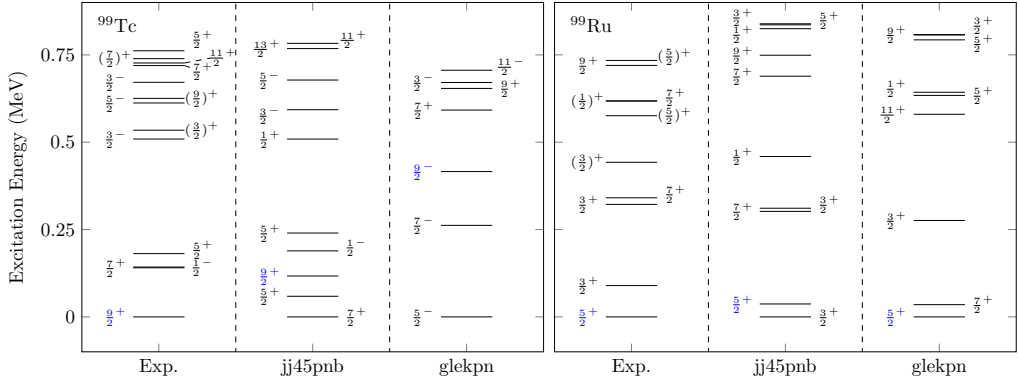

Fig. 1.: Energy level schemes for $^{99}$Tc and $^{99}$Ru with their corresponding NSM-computed values using the interactions *jj45pnb* and *glekpn*. Experimental data are taken from the ENDSF [21] evaluation. Levels in blue are the initial and final states relevant for this work.

With the considered truncations we have calculated the excitation spectra for both interactions and compare them with the experimental spectrum in Fig 1. It can be seen that the *jj45pnb* Hamiltonian works, overall, much better than the *glekpn* one, in particular for the ground state of $^{99}$Tc.

To further assess the reliability of the Hamiltonians employed, we have analyzed the magnetic dipole and electric quadrupole moments of the $^{99}$Tc and $^{99}$Ru ground states. The analysis can be seen in Table I, where once again the *jj45pnb* Hamiltonian does better overall than the *glekpn* Hamiltonian.

**2˙2.** *$\beta$ spectral shapes and half-lives*. – The half-life of a $\beta$ transition can be obtained from $t_{1/2} = \kappa/\tilde{C}$, where $\kappa$ is a constant [14, 15] and $\tilde{C}$ is the integrated shape function $C$,

---

($^1$) Orbital fixed with maximum amount of nucleons, e.g, 8 for $0f_{7/2}$.
($^2$) Orbital with at least 4 particles.

TABLE I.: Comparison of experimental and theoretical (*jj45pnb* and *glekpn* Hamiltonians) nuclear magnetic dipole (in units of nuclear magneton) and electric quadrupole (in units of barns) moments for $^{99}$Tc and $^{99}$Ru, for the experimental ground states and their correspondents in theory.

|  | $^{99}$Tc($9/2^+$) | | | $^{99}$Ru($5/2^+$) | | |
|---|---|---|---|---|---|---|
|  | Exp. | jj45pnb | glekpn | Exp. | jj45pnb | glekpn |
| $\mu$ ($\mu_N$) | +5.678(2) | +5.662 | +5.4160 | -0.641(5) | -0.2875 | +1.0826 |
| Q (b) | -0.129(6) | +0.0218 | +0.2428 | +0.079(4) | -0.0018 | -0.3222 |

TABLE II.: Values of the small relativistic NME emerging from the fitting procedure for the $^{99}$Tc$(9/2^+) \to$ Ru$(5/2^+)$ decay. The sNME values (1) and (2) shown reproduce the experimental half-life for their given $g_\text{A}^\text{eff}$.

| jj45pnb - CVC = −0.1705 | | | glekpn - CVC = −0.0080 | | |
|---|---|---|---|---|---|
| $g_\text{A}^\text{eff}$ | sNME$^{(1)}$ | sNME$^{(2)}$ | $g_\text{A}^\text{eff}$ | sNME$^{(1)}$ | sNME$^{(2)}$ |
| 0.8 | 0.0496 | -0.0856 | 0.8 | 0.1042 | -0.0311 |
| 0.9 | 0.0582 | -0.0772 | 0.9 | 0.1091 | -0.0257 |
| 1.0 | 0.0666 | -0.0687 | 1.0 | 0.1140 | -0.0202 |
| 1.1 | 0.0750 | -0.0601 | 1.1 | 0.1189 | -0.0148 |
| 1.2 | 0.0833 | -0.0515 | 1.2 | 0.1237 | -0.0093 |

full of phase-space factors and NMEs in the next-to-leading-order expansion, as discussed in detail in [9, 10]. Our calculations employ screening, radiative, and atomic exchange corrections. The most essential to the current study, due to its small $\beta$-decay $Q$ value, is the atomic exchange correction which was originally derived for allowed $\beta$ decays [22]; the effects of this correction are the most important for electron energies below 50 keV.

The complexity of the shape function $C$ can, however, be cast in a very simple dependence on the weak couplings by writing

$$C(w_e) = g_\text{V}^2 C_\text{V}(w_e) + g_\text{A}^2 C_\text{A}(w_e) + g_\text{V} g_\text{A} C_\text{VA}(w_e)\,, \tag{1}$$

where $w_e$ is the total (mass plus kinetic) energy of the emitted electron. The variation of the shape of the $\beta$-electron spectrum with the value of $g_\text{A}$ comes from the subtle interference of the combined vector $C_\text{V}(w_e)$ and axial $C_\text{A}(w_e)$ parts with the mixed vector-axial part $C_\text{VA}(w_e)$ [9].

For the currently studied 2nd-nu decay transition $^{99}$Tc$(9/2^+) \to {}^{99}$Ru$(5/2^+)$ we take the CVC-compatible value $g_\text{V} = 1.0$ for the weak vector coupling, and explore the impact of the quenched $g_\text{A}^\text{eff}$ values, ranging from $0.8 - 1.2$ (see Table II), on the $\beta$ spectral shape.

2˙3. *Small relativistic Nuclear Elements*. – An important contributor to the $\beta$ spectral shape and the half-life is the small relativistic NME (sNME) [12, 15, 16, 17]. In these works, by fitting the value of the sNME, alongside with $g_\text{A}^\text{eff}$, one is able to reproduce both the $\beta$ spectral shape and the half-life of a $\beta$ transition. The computation of the value of the sNME is a challenge for the ISM. This is why its CVC value (also presented in Table II) can be used as a reference [3]. It should be noted, though, that the CVC value represents an "ideal", pertaining to a perfect many-body theory, which is not the case for NSM, mainly due to its restricted proton and neutron valence spaces.

Following the philosophy adopted in [12, 15, 16, 17], we search for the values of the sNME that reproduce the experimental half-life of the $^{99}$Tc$(9/2^+) \to$ Ru$(5/2^+)$ transition. Since the $\beta$-decay rate has a quadratic dependency on the sNME, two solutions are usually found for each value of $g_\text{A}^\text{eff}$, as presented in Table II.

## 3. – Results for the $\beta$ spectral shapes

The $\beta$-decay transition of interest, $^{99}$Tc$(9/2^+) \to$ Ru$(5/2^+)$, is 2nd-nu and is sensitive to the $Q$ value of the decay. Thus, to ensure reliability of our $\beta$ spectral-shape calcula-tions, we have fixed the energy difference of the initial $9/2^+$ and final $5/2^+$ state by the experimental $Q$ value. We plot our results for the $\beta$ spectral shapes of this transition in Fig. 2. There the crossed-blue (circled-red) curves correspond to a typical moderately quenched value $g_A^{\text{eff}} = 1.0$ [4] and the fitted sNME value closer to (farther from) its CVC value, see Table II. The gray-hatched (gray-crossed) area, in turn, corresponds to the range $g_A^{\text{eff}} = 0.8 - 1.2$ with all the sNME values chosen closest to (farthest from) the corresponding CVC values, the sNME values being shown in Table II. Interestingly enough, the two different options for the sNME value (the two different hatched regions in Fig. 2) sometimes produce drastically different shapes, as can be seen for *glekpn*, a pattern also recorded in [17]. As for the interaction *jj45pnb*, both hatched regions overlap nicely and there seems not to be any strong dependency of the $\beta$ spectral shape on the choice of the value of sNME.

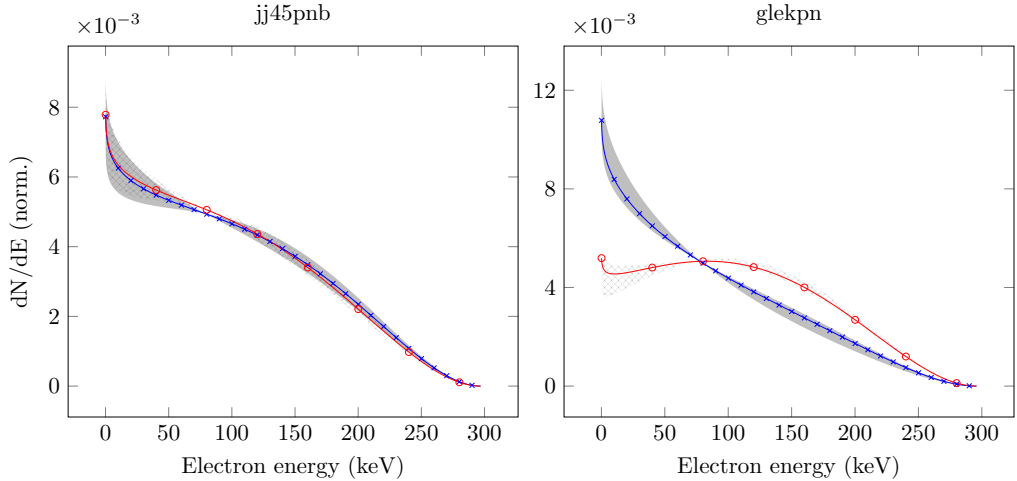

Fig. 2.: Computed $\beta$ spectral shapes for the decay transition $^{99}$Tc$(9/2^+) \to$ Ru$(5/2^+)$ using the Hamiltonians *jj45pnb* (left panel) and *glekpn* (right panel). The crossed-blue curves correspond to the fitted sNME closest to the CVC-value and the circled-red to the farthest (see Table II), with $g_A^{\text{eff}} = 1.0$. The gray-hatched (gray-crossed) area corresponds to the range $g_A^{\text{eff}} = 0.8 - 1.2$ with all the fitted sNME values chosen closest to (farthest from) the corresponding CVC values. All curves are normalized to unit area under them.

Since the $\beta$ spectral shapes produced by the two interactions clearly disagree, it is hard to judge which one of the shapes is the more realistic one. Based on both the level schemes in Fig. 1 and the magnetic and electric moments of Table I it is tempting to say that the *jj45pnb* Hamiltonian is better suited for these particular nuclei. Furthermore, looking at a recently measured experimental $\beta$ spectral shape for this transition in [23], especially in their Fig. 5, one can see a striking resemblance of it to the spectral shapes for both fitted sNME$^{(1,2)}$ of the *jj45pnb* Hamiltonian (see Table II) and a bit less striking for the solution sNME$^{(2)}$ for the *glekpn* Hamiltonian. However, only a proper comparison with data will allow us to draw any definite conclusions concerning the effective value of the axial coupling and the proper value of sNME. This we leave as a future exercise.

## 4. – Summary and conclusions

We have performed shell-model calculations of the $\beta$ spectral shape related to the second-forbidden non-unique $\beta^-$-decay transition $^{99}$Tc$(9/2^+) \to {}^{99}$Ru$(5/2^+)$, with a 100% branching ratio. We use a state-of-the-art $\beta$-decay formalism and compute the many involved NME using the well established NSM Hamiltonians *jj45pnb* and *glekpn*. When compared with future available data, the $\beta$-electron spectrum of this transition is potentially indicative of the effective value of the weak axial coupling, $g_A$, of crucial importance for extraction of information on beyond-the-standard-model physics from the results of the present and future rare-events experiments. We also point out the important role played by the small relativistic vector matrix element, used to reproduce the measured half-life of the transition, in shaping the $\beta$-electron spectrum.

## 5. – Acknowledgments

We acknowledge the support by CSC – IT Center for Science, Finland, for the generous computational resources, which were essential to this project.